\documentclass[prl, twocolumn]{revtex4-2}

\usepackage{amssymb,amsmath}
\usepackage{latexsym}
\usepackage{graphicx}
\usepackage{color}
\usepackage{bm}      
\usepackage{float} 
\usepackage[colorlinks=true, linkcolor=blue, citecolor=blue, urlcolor=blue]{hyperref}  

\usepackage{dcolumn}

\def\ud{\mathrm{d}}

\DeclareFontFamily{U}{mathx}{}
\DeclareFontShape{U}{mathx}{m}{n}{<-> mathx10}{}
\DeclareSymbolFont{mathx}{U}{mathx}{m}{n}
\DeclareMathAccent{\widehat}{0}{mathx}{"70}
\DeclareMathAccent{\widecheck}{0}{mathx}{"71}

\begin{document}
\title{Finite-resolution measurement induces topological curvature defects in spacetime}

\author{Ewa Czuchry}
\email{ewa.czuchry@wat.edu.pl} \affiliation{Institute of Mathematics and Cryptology, Military University of Technology, ul. gen. Sylwestra Kaliskiego 2,
00-908 Warszawa 46, Poland}
\author{Jean-Pierre Gazeau}
\email{gazeau@apc.in2p3.fr}
\email{j.gazeau@uwb.edu.pl}
\affiliation{Universit\'e  Paris Cit\'e, CNRS, Astroparticule et Cosmologie, F-75013 Paris, France}
\affiliation{Faculty of Mathematics, University  of Bia\l ystok, 15-245 Bia\l ystok, Poland}

\begin{abstract}
We show that regularizing  $(2+1)$-dimensional Minkowski spacetime  with a finite-resolution Gaussian probe, analogous to Weyl-Heisenberg (Gabor) signal analysis and related quantization, induces a curved geometry with a topological defect. The regularized metric replaces $r^2$ by $r^2+\sigma^2$ in the angular part, where $\sigma$ is the resolution scale from the width of the Gaussian probe. The resulting Gaussian curvature integrates to $-2\pi$, independently of $\sigma$. This curvature defines an effective stress-energy source with universal total energy $E_{\text{eff}}=-1/(4G)$.  The limit $\sigma\to0$  leads to  distributional Dirac-delta curvature and to appearance of topological defect at the origin. These results show that finite spatial resolution measurement does not merely smooth singularities but can shape spacetime geometry. 
\end{abstract}

\maketitle

\noindent{\it Introduction.} Spacetime singularities in General Relativity create a problem to  our understanding of physics. While curvature singularities like the Big Bang represent explicit geometric pathologies, coordinate singularities, such as those at the origin in polar coordinates or at event horizons, arise from mathematical description rather than physical reality. These coordinate artifacts typically rely on an unphysical idealization: the ability to localize spacetime points with infinite resolution.

In reality, any measurement of geometry is mediated by finite-resolution probes, making exact point-localization physically inaccessible. This suggests that coordinate singularities may reflect an over-idealized mathematical description rather than intrinsic features of spacetime. A consistent treatment of finite resolution should therefore provide a physically motivated regularization of such singularities. This naturally raises the question: how does finite-resolution probing affect spacetime geometry, even in flat backgrounds? Can coordinate singularities be regularized in a way consistent with the operational limitations of measurement?

In this work, we address this question by applying a Weyl–Heisenberg (Gabor) regularization scheme directly to the spacetime metric. This approach provides a natural framework to implement finite-resolution effects via Gaussian smearing with width $\sigma$. Applied to flat Minkowski spacetime metrics written in polar coordinates, it replaces the term $r^2$ with $r^2+\sigma^2$, removing the coordinate singularity at the origin. We show that this procedure induces genuinely new geometry, not obtainable by smooth coordinate transformations. The resulting Gaussian curvature $K(r)=-\sigma^2/(r^2+\sigma^2)^2$ integrates to $-2\pi$, independent of $\sigma$. In the limit $\sigma\to0$, the curvature localizes into a delta-function and provides  Euler characteristic of the spatial part $\chi=0$, corresponding to the punctured plane $\mathbb{R}^2\setminus \{0\}$ of the spatial part. This way a  coordinate singularity has been converted  into a topological defect. The geometry carries an effective stress-energy with total energy $E_{\mathrm{eff}}=-1/(4G)$, universal and independent of $\sigma$, indicating a fundamental gravitational cost associated with localization.\\

\noindent{\it Weyl--Heisenberg regularisation of Minkowski spacetime.} To establish these results, we now introduce the Weyl--Heisenberg regularization scheme and its implementation for spacetime metrics. The method employed in this work comes from our previous papers on the Weyl--Heisenberg (coherent-state) quantisation \cite{Bergeron:2018und} of classical phase-space functions, which in signal analysis corresponds to the {so-called  Gabor (windowed Fourier) quantization} \cite{cogahasha22}. This approach maps functions on phase space to operators via {smooth  smearing, e.g., Gaussian}, implementing finite resolution in both position and momentum {or wave-vector}.

{As a one-dimensional illustration, the Gabor quantization of a function $u(x)$
with a Gaussian probe
$$
G_{\sigma}(x)=\frac{1}{\pi^{1/4}\sqrt{\sigma}}
\,\exp\!\left(-\frac{x^{2}}{2\sigma^{2}}\right)
$$
yields its semiclassical portrait as the convolution
$$
\widecheck{u}(x)=\bigl(u\ast G^{2}_{\sqrt{2}\sigma}\bigr)(x)\,.
$$
Applied to a spacetime metric field $g_{\mu\nu}(x)$, this prescription amounts
to a convolution with a rotationally invariant kernel of width~$\sigma$,
\begin{equation}
\widecheck{g}_{\mu\nu}(x)
=\int \mathrm{d}^{4}x'\,g_{\mu\nu}(x')\,K_{\sigma}(x-x')\,,
\end{equation}
where $K_{\sigma}$ is normalized and rapidly decaying beyond the scale
$\sigma$. This procedure preserves rotational symmetries while introducing a
minimal resolvable length.}

{When applied to $(2+1)$-dimensional Minkowski spacetime, the convolution
leaves the constant metric components unchanged in Cartesian coordinates.
In polar coordinates $(r,\theta)$, where the flat line element reads
$$
\ud s^{2}=-\ud t^{2}+\ud r^{2}+r^{2}\ud\theta^{2}\, , 
$$
only the angular component $g_{\theta\theta}=r^{2}$ is non constant and therefore
affected by the smearing. Restricting to the two-dimensional spatial sector and
using a Gaussian kernel
$$
K_{\sigma}(\mathbf{ x})=\frac{1}{2\pi\sigma^{2}}
\,\exp\!\left(-\frac{|\mathbf{ x}|^{2}}{2\sigma^{2}}\right)\,,
$$
the convolution yields
$$
(r^{2}\ast K_{\sigma})(r)=r^{2}+2\sigma^{2}\,.
$$}
Absorbing the numerical factor into a redefinition $\sigma^2 \to 2\sigma^2$, we obtain the regularised metric
\begin{equation}\label{MR}
\ud s^2=-\ud t^2+\ud r^2+(r^2+\sigma^2)\ud\theta^2 {= -\ud t^{2}+ \ud \tilde{s}^{2}\,.}
\end{equation}
where  $\ud \tilde{s}^{2}$ denotes the spatial part:
\begin{equation}
\label{smet}
\ud \tilde{s}^2 = \ud r^2 + (r^2+\sigma^2)\,\ud\theta^2\, .
\end{equation}
The parameter $\sigma$ thus controls the resolution  scale, smearing the coordinate singularity at $r=0$ while maintaining the asymptotic flat structure. \\

\noindent{\it Geometric analysis of the regularized metric.} For $\sigma>0$  the geometry described by the metric \eqref{MR} is no longer flat.  Its curvature can be computed directly using standard Riemannian geometry. The non-zero Christoffel symbols are:
\begin{align}
\Gamma^\theta_{r\theta} = \Gamma^\theta_{\theta r} = \frac{r}{r^2+\sigma^2}, \quad
\Gamma^r_{\theta\theta} = -r\, .
\end{align}
Subsequently,  the Riemann tensor components $R^\rho_{\ \sigma\mu\nu}$ are obtained  and  can be contracted  to the Ricci tensor:
$$
R_{\mu\nu} = R^\rho_{\ \mu\rho\nu}\,,
$$
with non-zero components
\begin{equation}\label{Ricci}
R_{rr} = -\frac{\sigma^2}{(r^2+\sigma^2)^2},\quad  R_{\theta\theta} = -\frac{\sigma^2}{r^2+\sigma^2},
\end{equation}
and the Ricci scalar
\begin{equation}\label{scalar}
R = g^{\mu\nu}R_{\mu\nu} = -\frac{2\sigma^2}{(r^2+\sigma^2)^2}\,.
\end{equation}
The regularization therefore curve the metric and is not  a mere coordinate artifact.
\begin{figure}[htb!]
\centering
\includegraphics[width=0.5\textwidth]{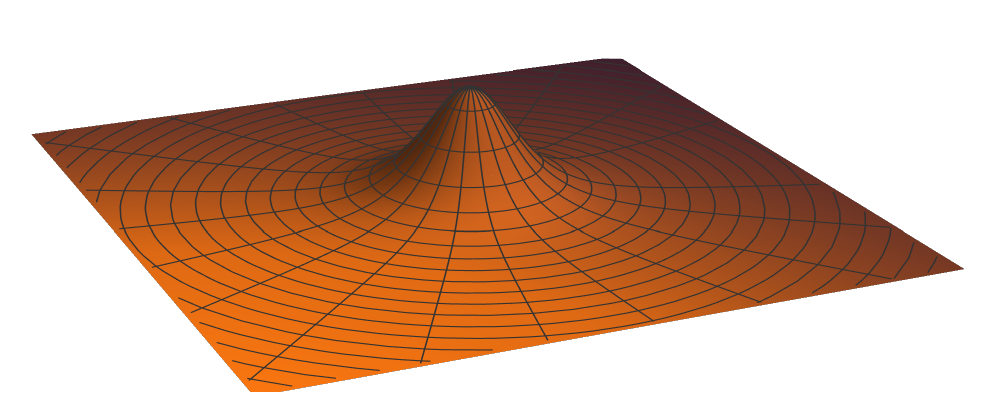}
\caption{Curved geometry.}
\label{fig:curved}
\end{figure}\\

\begin{figure}[htb!]
\centering
\includegraphics[width=0.3\textwidth]{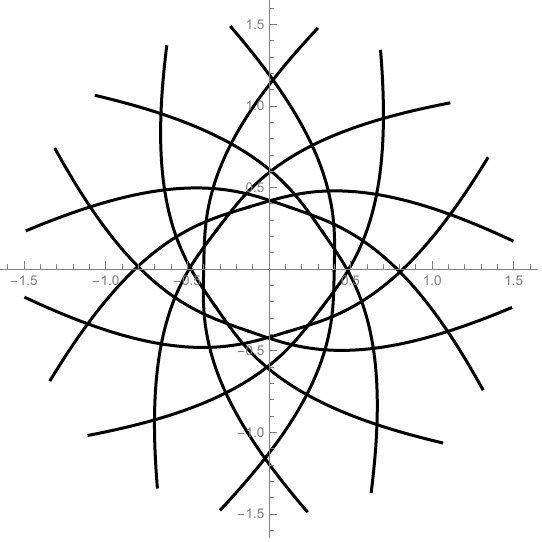}
\includegraphics[width=0.5\textwidth]{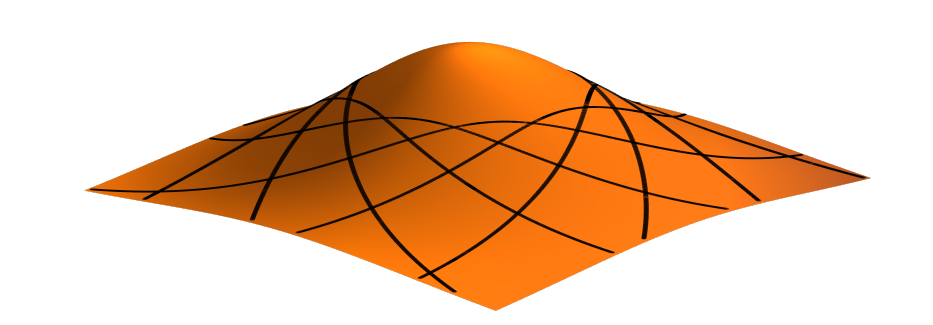}
\caption{Sample geodesics for  $\sigma=1$ on 2D space and their visualization on a curved surface. }
\label{fig:geodesics}
\end{figure}

\noindent{\it Spatial geometry as a minimal helicoid.} The $t=\text{const}$ spatial slice of the regularized metric \eqref{MR} can be isometrically embedded in $\mathbb{R}^3$. Starting from the  2D spatial metric \eqref{smet}
we seek an embedding $(X,Y,Z)$ into Euclidean space with metric $\ud S^2= \ud X^2+\ud Y^2+\ud Z^2$.

Identifying $X=r\cos\theta$, $Y=r\sin\theta$, the embedding conditions force $Z$ to depend only on $\theta$, with $(\partial_\theta Z)^2 = \sigma^2$. The solution yields the helicoid equation with non-standard parametrization:
\begin{equation}\label{helicoid}
X = r\cos\theta,\quad Y = r\sin\theta,\quad Z = \pm \sigma\theta.
\end{equation}
The alternative $Z = -\sigma\theta$ gives the mirror image with opposite helicity.

\begin{figure}[htb!]
\centering
\includegraphics[width=0.4\textwidth]{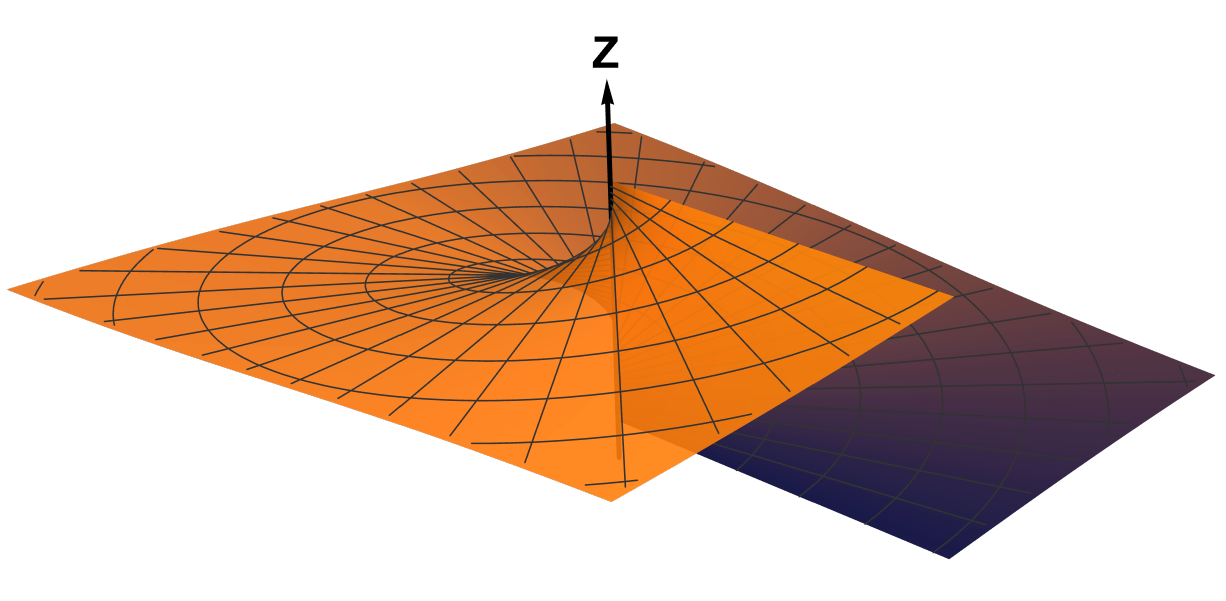}
\caption{Helicoidal surface $Z = \sigma\phi$ with a pitch $2\pi\sigma$.}
\label{fig:helicoid}
\end{figure}

The helicoid is a minimal surface (mean curvature $H=0$) whose Gaussian curvature $K(r)$ matches exactly that computed intrinsically:
\begin{equation}\label{GC}
K(r) = -\frac{\sigma^2}{(r^2+\sigma^2)^2}\,.
\end{equation}
Lines of constant $r$ are helices of pitch $2\pi\sigma$, making $\sigma$ the geometric pitch parameter.

As $\sigma \to 0$, the embedding collapses to the plane $Z=0$ for $r>0$, but the line $r=0$ maps to a vertical segment of length $2\pi\sigma$ that shrinks to a point only in the strict limit. \\

\noindent{\it Topology} The metric   \eqref{MR}  defines a smooth two-dimensional Riemannian manifold whose topology is
that of a half-cylinder, $\mathbb{R}_{\geq 0} \times \mathbb{S}^1 $. The periodic identification of the angular coordinate implies a non-contractible circular direction, while the radial coordinate $r \in \mathbb{R}_{\geq 0}$
extends over the half-line. Since $g_{\theta\theta}=r^2+\sigma^2$
never vanishes for $\sigma>0$, the orbit $r=0$ is not a point but a circle
of minimal circumference $2\pi\sigma$.
In the limit $\sigma \to 0$, the minimal
circle  shrinks in size  but remains
topologically a circle, which at $r=0$ becomes singular and the geometry  degenerates to that of punctured the plane $\mathbb{R}^2\setminus\{0\}$.

{Moreover, performing the coordinate transformation}
\begin{equation}
r = \sigma \sinh u ,
\end{equation}
{brings the metric \eqref{MR} to the conformal form}
\begin{equation}\label{conformal}
\ud {\bar{s}}^2 = \sigma^2 \cosh^2 u \left( \ud u^2 + \ud \theta^2 \right),
\qquad \theta \sim \theta + 2\pi .
\end{equation}

{This metric can be visualized as a curved surface embedded in $\mathbb{R}^3$ through the parametrization}
\begin{equation}\label{catenoid}
X=\sigma \cosh u \cos\theta,\quad 
Y=\sigma \cosh u \sin\theta,\quad 
Z=\pm \sigma u .
\end{equation}

\begin{figure}[htb!]
\centering
\includegraphics[width=0.3\textwidth]{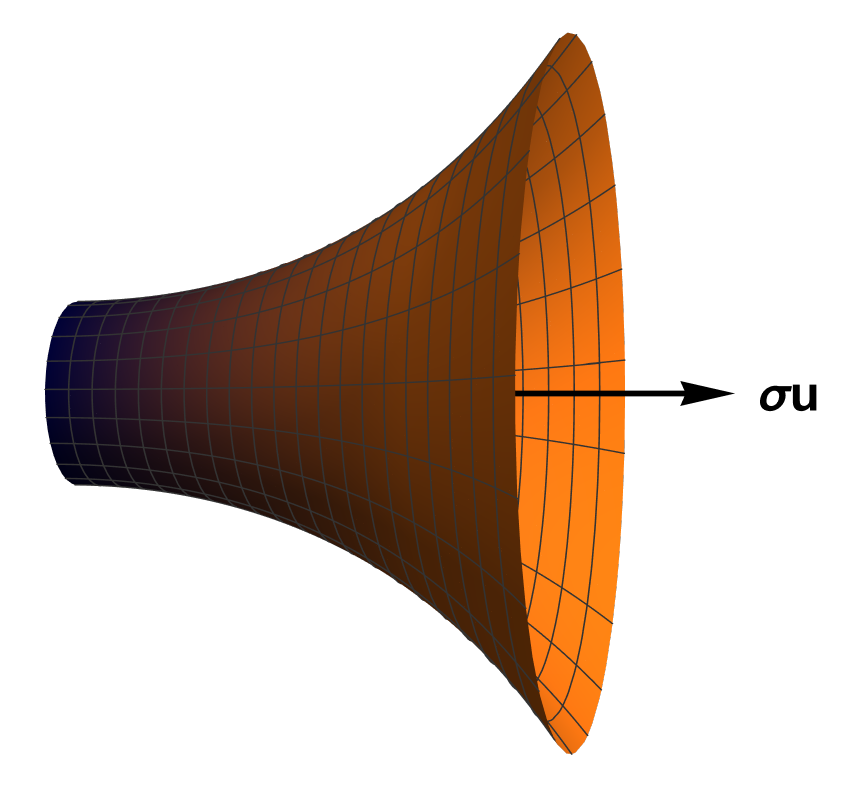}
\caption{Catenoid representation of the conformal geometry \eqref{conformal}.}
\label{fig:conformal}
\end{figure}

{These equations describe the standard embedding of a catenoid, which is also related to the helicoid through a well-known family of minimal surfaces. The geometry is therefore conformally equivalent to a flat half-cylinder with conformal factor $\sigma^2\cosh^2 u$. At $u=0$ the catenoid reaches its minimal radius, which is finite and equal to $\sigma$, giving a minimal circumference $2\pi\sigma$. This representation makes the topology manifest, as illustrated in Fig.~\ref{fig:conformal}.}
\\

{If one analytically continues the coordinate $u$ to negative values, the manifold naturally extends to a full cylinder with topology $\mathbb{R}\times \mathbb{S}^1$. The surface then possesses a minimal but non-zero throat at $u=0$, connecting two asymptotically cylindrical regions corresponding to $u\to\pm\infty$. In this extended description the geometry is geodesically complete, since the geodesic equations can be smoothly continued across the throat into the region with negative values of $u$ (or equivalently negative $r$ in the analytic extension). The resulting structure resembles the Einstein-Rosen bridge \cite{einros35}, although its physical interpretation is different. In particular, in $(2+1)$-dimensional gravity there is no corresponding black-hole solution in the present setting, and the geometry should instead be understood as a smooth minimal-surface bridge.}\\
\begin{figure}[htb!]
\centering
\includegraphics[width=0.5\textwidth]{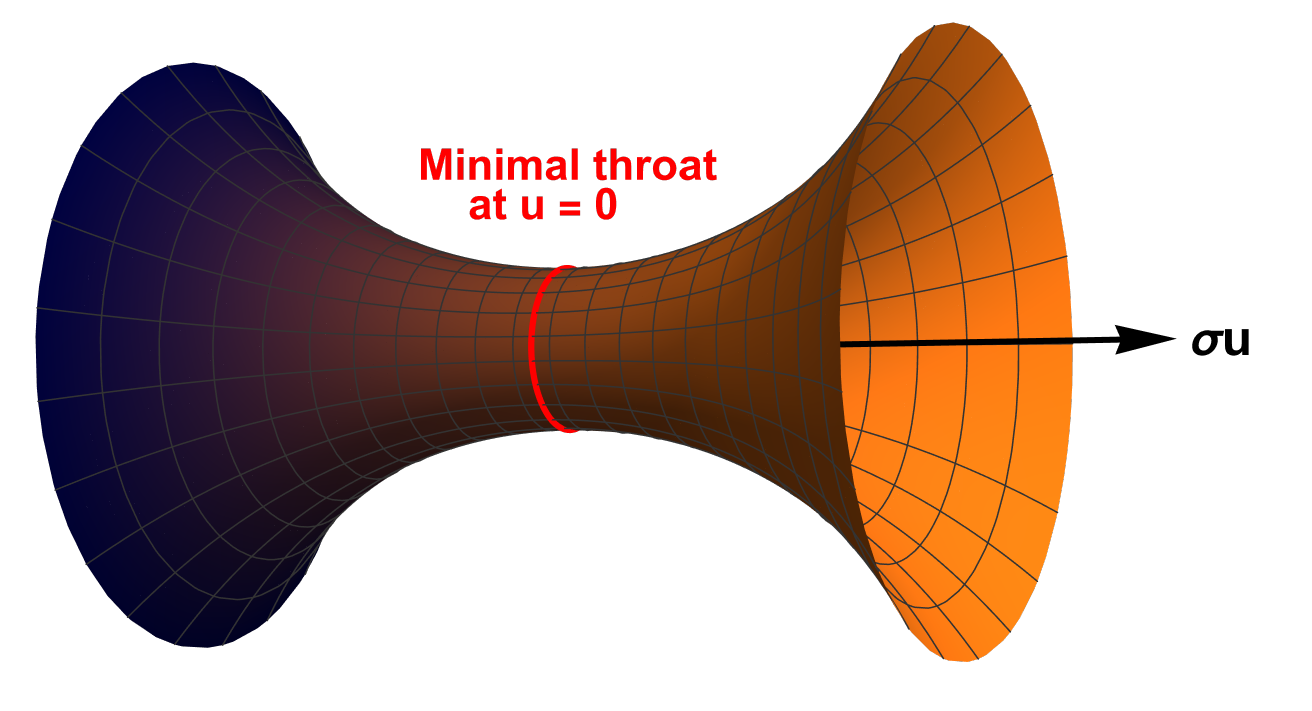}
\caption{Analytical extension of the conformal solution.}
\label{fig:extension}
\end{figure}\\

\noindent{\it Measurement-induced curvature and topological charge.}
The Gaussian curvature $K(r)$ of the metrics \eqref{MR} is localized near the origin and depends on the value of $\sigma$.Its integral over the spatial slice is constant and independent of $\sigma$:
\[
\int_{\mathbb{R}^2} K(\mathbf{x})\,\ud A
= \int_{\mathbb{R}^2} \frac{R}{2}\,\ud A = -2\pi,
\]
where $\mathbf{x}\in \mathbb{R}^2$, $\ud A=\sqrt{r^2+\sigma^2}\,\ud r \ud\theta$, with $r = |\mathbf{x}|$, 
and the curvature scalar is given by \eqref{scalar}.

Moreover, in the limit $\sigma\to0$, one has $K(\mathbf{x})\to\mathbf{0}$ pointwise for $\mathbf{x}\neq\mathbf{0}$, while 
\[
K(\mathbf{0})=-\frac{1}{\sigma^2}\xrightarrow{\;\sigma\to0\;} -\infty.
\]
Indeed, for any smooth test function $f(\mathbf{x})$,
\[
\lim_{\sigma\to0}\int_{\mathbb{R}^2} K(\mathbf{x})\,f(\mathbf{x})\,\ud A 
= -2\pi f(\mathbf{0}),
\]
so that
\begin{equation}\label{DistribK}
K(\mathbf{x})\xrightarrow{\mathcal{D}'} -2\pi\delta^{(2)}(\mathbf{x})\,.
\end{equation}
Therefore, at $r=|\mathbf{x}|=0$ there is  a topological curvature defect.

The topological meaning of this defect becomes clear when applying the Gauss--Bonnet theorem to the curvature in the distributional limit. Let us integrate the   curvature \eqref{DistribK} over a large disk of radius $R$ containing the origin plus over its boundary. The boundary is a circle in the flat metric with geodesic curvature $k_g=1/R$. Thus 
\[
\int_{\text{disk}} K\,\ud A + \int_{\partial\text{disk}} k_g\,\ud s = (-2\pi) + 2\pi = 0.
\]
According to Gauss--Bonnet theorem, this equals $2\pi\chi$, giving Euler characteristic $\chi=0$. It is exactly the same as  that of a punctured plane $\mathbb{R}^2\setminus\{0\}$.

Curvature concentrated on lower-dimensional sets can be assigned an invariant ``defect charge'' $\delta=\int K \ud A = -2\pi $, as discussed in the context of conical singularities and $(2+1)$ gravity \cite{Visser90,GerochTraschen87,Deser}, and 
in terms of distributional curvature measures \cite{Penrose72,doCarmo76}. In view of conical singularities approach, the integrated curvature corresponds to an excess angle $2\pi$.  However, the rotational holonomy corresponding to that value is trivial, and the system therefore does not correspond to an ordinary conical geometry with deficit angle $\delta\in(0,2\pi)$. 

\begin{figure}[htb!]
\centering
\includegraphics[width=0.5\textwidth]{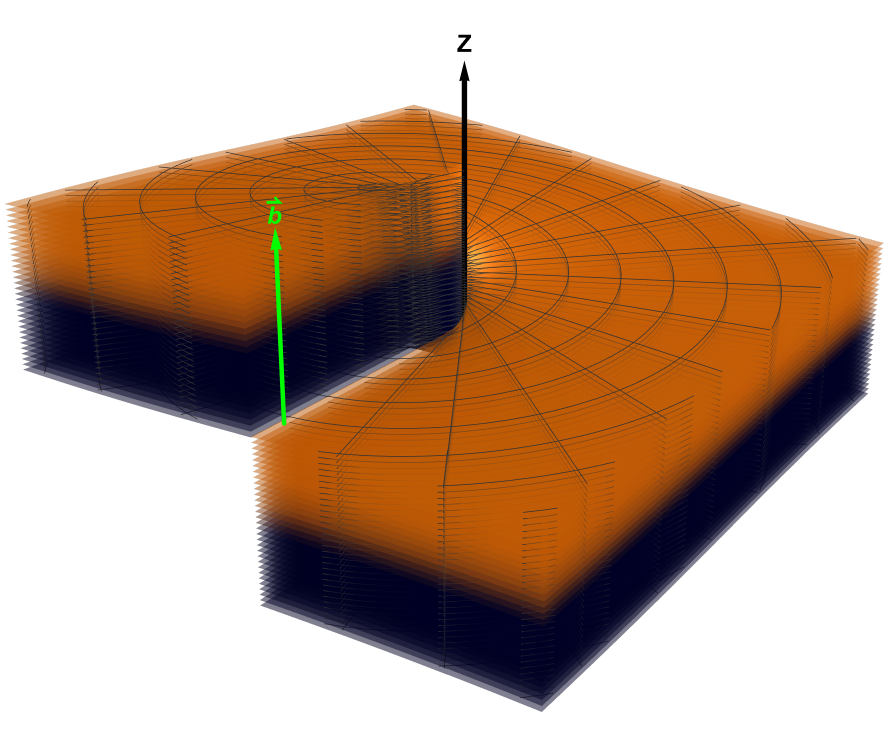}
\caption{Line screw defect with a Burgers vector $\mathbf{b}=[0,0,2\pi\sigma]$.}
\label{fig:mhelicoid}
\end{figure}

This topological defect inherent to $(2+1)$ gravity  can be also described in terms of defects theory developed for crystals and $(2+1)$-dimensional gravity \cite{Katanaev:1992kh}. Namely, it corresponds to full disclination created by  inserting a wedge of angle $2\pi$.
Such a $2\pi$ disclination  is topologically trivial in rotation, therefore it  is equivalent to a dislocation with a Burgers vector $\mathbf{b}$ 
\cite{Katanaev:1992kh}. The  defect is therefore a pure screw dislocation,  on a helicoid embedding the Burgers vector  $\mathbf{b}=[0,0,2\pi \sigma]$   describes a pitch between helicoid leaves. It is shown in figure \eqref{fig:mhelicoid} for $(3+1)$D cylindrical case.\\

\noindent{\it Effective stress–energy and defect energy.} From the Ricci tensor \eqref{Ricci} and scalar \eqref{scalar}
the Einstein tensor $G_{\mu\nu} = R_{\mu\nu} - \frac{1}{2}R g_{\mu\nu}$ has components
\begin{equation}
G_{00} = - \frac{\sigma^2}{(r^2+\sigma^2)^2}\,, \quad
G_{rr} = G_{\theta\theta} = 0\,.
\end{equation}
Interpreting this through Einstein's equations provides an effective energy-momentum tensor characterized by a localized, negative energy density
\begin{equation} \label{eq:eff_energy_density}
\rho_{\mathrm{eff}}(r) = - \frac{\sigma^2}{(r^2+\sigma^2)^2}\,.
\end{equation}
The density peaks at the origin with $\rho_{\mathrm{eff}}(0) = - 1/\sigma^2$ and decays as $1/r^4$ for $r \gg \sigma$.

It turns out that the total integrated effective energy is independent of the regularization scale:
\begin{equation} \label{eq:total_energy}
E_{\mathrm{eff}} = \int \rho_{\mathrm{eff}} \sqrt{-g}\, \ud^2x = -2\pi\,,
\end{equation}
where space-time volume element $\sqrt{-g} = \sqrt{r^2+\sigma^2}$. Restoring dimensions (with $8\pi G=1$),
\begin{equation}
E_{\mathrm{phys}} = -\frac{1}{4G} \sim -M_P^2\,,
\end{equation}
with $M_P = G^{-1/2}$ being the Planck mass. The total energy is thus fixed only  by gravitational constant $G$.

The constant total energy  resembles the constant topological charge. Indeed, the same distributional analysis as for curvature scalar applies to the effective energy density $\rho_{\mathrm{eff}}(|\mathbf{x}|)$ \eqref{eq:eff_energy_density}:
\begin{equation}
\rho_{\mathrm{eff}}(|\mathbf{x}|) \xrightarrow{\mathcal{D}'} -\frac1{4G}\delta^{(2)}(\mathbf{x})\,,
\end{equation}
This corresponds to a point particle of negative mass $M = -1/(4G)$. This matches the known result that in $(2+1)$-dimensional gravity point masses are described by conical defects \cite{Staruszkiewicz,Deser} with deficit angle $\delta = 8\pi GM$; here $\delta = -2\pi$ gives also  $M = -1/(4G)$.\\

\noindent{\it Extension to $(3+1)$ dimensions and dimensional comparison.} The Gabor regularization scheme naturally extends to higher dimensions. While the $(2+1)$D case regularizes a point singularity in polar coordinates, its natural counterpart in $(3+1)$D is the line singularity of cylindrical coordinates $(t,r,\phi,z)$. Regularizing the flat metric $\ud s_c^2 = -\ud t^2 + \ud r^2 + r^2 \ud\phi^2 + \ud z^2$ 
\cite{cogahasha22} yields
$$
\ud s^2 = -\ud t^2 + \ud r^2 + (r^2+\sigma^2)\ud\phi^2 + \ud z^2 \, .
$$

The curvature components are non-zero, with Ricci scalar
$$
R_c = -\frac{2\sigma^2}{(r^2+\sigma^2)^2}\,,
$$
and Einstein tensor
$$
G_{\mu\nu}^{(c)}= \text{diag}\!\left(-\frac{\sigma^2}{(r^2+\sigma^2)^2},\,0,\,0,\,\frac{\sigma^2}{(r^2+\sigma^2)^2}\right)\,.
$$

As $\sigma\to0$, the curvature localizes distributionally to the $z$-axis:
$$
R_{c} \xrightarrow{\mathcal{D}'} -4\pi\,\delta^{(2)}(x,y)\,,
$$
leading to distribution-like Einstein tensor component:
$$
G_{00}^{(c)} \xrightarrow{\mathcal{D}'} -2\pi\,\delta^{(2)}(x,y), \qquad
G_{33}^{(c)} \xrightarrow{\mathcal{D}'} +2\pi\,\delta^{(2)}(x,y)\,.
$$
This corresponds to  line defect with stress-energy
$$
T_{\mu\nu}^{(c)} =- \frac{1}{4G}\,\delta^{(2)}(x,y)\,\text{diag}(1,0,0,-1)\,,
$$
with  negative energy density $\rho_c = -1/(4G)$ per unit length, and positive tension along the axis. The geometry around the defect exhibits an excess angle ($\delta_c < 0$) unlike  than the deficit angle of a standard cosmic string.

By contrast, regularizing spherical coordinates $(t,r,\theta,\phi)$ in $(3+1)$ dimensions \cite{cogahasha22}  provides the regularised metrics in the form:
$$
ds_s^2=-\ud {x_0}^2 + \ud r^2 + \left( r^2 + 3\sigma^2 \right)\ud \theta^2 + (\rho^2 + 2\sigma^2)\ud \phi^2\,.
$$
where $\rho^2=r^2\sin^2\theta$. Here the singularity is a point, with  qualitatively different behavior provided by preliminary calculations: the curvature does not localize to a delta function but  forms an extended anisotropic fluid halo with $R_c\sim -2/r^2$ as $\sigma\to0$, leading to a divergent total effective energy. 
Describing such defects in three spatial dimensions requires moving beyond the Gauss-Bonnet theorem, which applies only to 2D surfaces. \\

\noindent{\it Summary and conclusions.} We have shown that finite-resolution measurement, implemented via Weyl-Heisenberg (Gabor) regularization fundamentally alters spacetime geometry. The probing Gaussian probe introduces regularisation scale $\sigma$. Its application to $(2+1)$ Minkowski in singular polar coordinates smears the coordinate singularity while preserving rotational symmetry and symmetry-adapted coordinates. 
  
One would expect that the limit $\sigma\to 0$  recovers the original unprobed flat metric. However, we demonstrated that in this limit curvature scalar acquires  distributional form $K(\mathbf{x})\rightarrow-2\pi\delta^{(2)}(\mathbf{x})$, creating a topological puncture with Euler characteristics $\chi=0$. The effective energy density also acquires distributional form $\rho_{\mathrm{eff}}(r) \xrightarrow{\mathcal{D}'} -2\pi\delta^{(2)}(\mathbf{x})\,$.  In $(3+1)$ dimensions we have obtained  a line defect with energy per unit length $\mu = -1/(4G)$, and stress-energy $T_{\mu\nu} = \frac{1}{4G}\delta^{(2)}(x,y)\text{diag}(-1,0,0,1)$. Interestingly, both cases share the same energy scale $1/(4G)$ and the same exhibit excess-angle value. However, the angle  $\delta =-2\pi$  does not produce an ordinary conical geometry, but  it manifests as  the gravitational analogue of a screw dislocation, consistent with the helicoidal embedding of spatial slices.

The described regularization scheme  exhibit  a link between geometry and measurement in general relativity. In the conventional approach  spacetime geometry is considered as a real object  existing independently of how it is probed. Here we demonstrate that any finite-resolution measurement necessarily changes  the metric. The observed  space-time becomes curved with a topological defect at the origin and the measurement itself results in energy release $E_{\mathrm{eff}}=-1/(4G)$ independent of the resolution scale and fixed solely by gravitational constant $G$. 

This energy can be interpreted as the cost of localization. Selecting a specific origin $r=0$ out of the continuum of equivalent points breaks translational symmetry and effectively introduces a topological defect. The associated curvature and energy arise as a necessary consequence of this symmetry breaking. In this sense, the procedure of measurement does not merely reveal geometry, but actively participates in its formation. 

An additional perspective is provided by information theory \cite{Cover}. For a Gaussian probe  centered at a location \(\bm{\theta}\in\mathbb{R}^{2}\):
\[
K_{\sigma}(\mathbf{x};\bm{\theta}) = \frac{1}{2\pi\sigma^{2}}\, 
e^{-\frac{|\mathbf{x}-\bm{\theta}|^{2}}{2\sigma^{2}}},
\]
The Fisher information matrix is independent of \(\bm{\theta}\) and diagonal: \( I_{ij}(\sigma) = \frac{1}{\sigma^{2}}\,\delta_{ij}.\) The corresponding \emph{Fisher information density} is the integrand evaluated at a fixed \(\bm{\theta}\), say \(\bm{\theta}=0\):
\begin{equation}
i_{F}(\mathbf{x};\sigma)    = \frac{|\mathbf{x}|^{2}}{2\pi\sigma^{6}}\,e^{-|\mathbf{x}|^{2}/(2\sigma^{2})}.
\label{eq:fisher-density}
\end{equation}
However, its integral over the plane that reproduces the trace 
is $\sigma$-dependent and divergent in the limit $\sigma\to 0$: \(\int_{\mathbb{R}^{2}} i_{F}(\mathbf{x};\sigma)\,\ud^{2}x = \frac{2}{\sigma^{2}}\).  On the other hand, the rescaled density $
j_{F}(\mathbf{x};\sigma) := \sigma^{2}\, i_{F}(\mathbf{x};\sigma)$ gives regular total localization  information. Also, in the distributional limit $j_{F}(\mathbf{x};\sigma)\;\xrightarrow{\mathcal{D}'}\;2\,\delta^{(2)}(\mathbf{x})$
and the total Fisher information, after rescaling, converges to
\[
\lim_{\sigma\to0}\int j_{F}\,\ud^{2}x = \int 2\delta^{(2)}(\mathbf{x})\,\ud^{2}x = 2.\]
As \(\sigma^{2}\) is proportional to the area of a resolution cell (the effective area of the Gaussian kernel), \(j_F\) represents the Fisher information contained in one such cell at position \(\mathbf{x}\). This rescaling isolates the ``finite informational content'' of the measurement process: the constant value $2$ can be interpreted as the total information content associated with fixing a point in two dimensions.
This shows that, in the limit of perfect resolution, all positional information becomes localized at the chosen origin. 

Taken together, these results suggest a possible correspondence between localization, curvature, and energy: concentrating information at a point appears to be accompanied by the appearance of a geometric defect with a fixed energy. In this picture, singularities may be viewed as limits of finite-resolution measurements, with the associated energy reflecting the cost of localization. A general framework relating energy, defects, and information, however, remains to be developed beyond black hole settings in $(3+1)$ gravity. 

More detailed calculations are presented in the accompanying longer work \cite{CzG}. In particular, while curvature tensors are derived here using Christoffel symbols, the punctured topology $\mathbb{R}^2\setminus\{0\}$ motivates a complementary anholonomic dreibein formulation. Both approaches provide the same results, with the latter providing a description better adapted to the global structure of the manifold. The extended analysis also includes the geodesic equations and their explicit solutions in terms of Jacobi elliptic functions.\\

 \noindent{\it Future prospects.}  Generally, $(2+1)$-dimensional gravity has well-known limitations: in particular, it lacks local propagating gravitational degrees of freedom, and its dynamics is entirely governed by global, topological features. As a result, the description of localized defects within this framework is necessarily indirect and relies on global constructions.

A more direct and geometrically transparent description of such defects is provided by the language of defect theory, originally developed in elasticity. This framework has been extended to spacetime in the context of teleparallel gravity \cite{Kleinert:1989ky, Kleinert:2008zzb}, where topological defects are described in terms of torsion rather than curvature. In this formulation, screw dislocations arise naturally as torsional defects, with the torsion tensor playing the role of a defect density. This perspective is complementary to the curvature-based description adopted in the present work and suggests an alternative interpretation of the localized structures we obtain.We have constructed a teleparallel counterpart of the regularized $(2+1)$ Minkowski spacetime, which will be presented soon. Further work extending these ideas, including the $(3+1)$-dimensional case, is currently in progress.\\

\noindent{\it Acknowledgments.} This work has been co-financed by Military University of Technology  under research project UGB 531-000089-W500-22.

\end{document}